\documentclass[12pt,preprint]{aastex}

\slugcomment{To appear in The Astrophysical Journal Letters}

\shorttitle{First HCO$^{+}$ Survey in LIRGs and ULIRGs}

\shortauthors{Graci\'{a}-Carpio et al.}

\begin{document}


\title{Is HCN a True Tracer of Dense Molecular Gas in LIRGs and ULIRGs?}

\author{J. Graci\'{a}-Carpio\altaffilmark{1}, S. Garc{\'{\i}}a-Burillo\altaffilmark{1}, P. Planesas\altaffilmark{1} and L. Colina\altaffilmark{2}}

\altaffiltext{1}{OAN, Alfonso XII 3, 28014 Madrid, Spain; j.gracia@oan.es, s.gburillo@oan.es, p.planesas@oan.es}

\altaffiltext{2}{IEM, CSIC, Serrano 121, 28006 Madrid, Spain; colina@damir.iem.csic.es}



\begin{abstract}
We present the results of the first HCO$^{+}$ survey probing the dense molecular gas content of a sample of 16 luminous and ultraluminous infrared galaxies (LIRGs and ULIRGs). Previous work, based on HCN(1--0) observations, had shown that LIRGs and ULIRGs posses a significantly higher fraction of dense molecular gas compared to {\it normal} galaxies. While the picture issued from HCO$^{+}$ partly confirms this result, we have discovered an intriguing correlation between the HCN(1--0)/HCO$^{+}$(1--0) luminosity ratio and the IR luminosity of the galaxy ($L_{\rm{IR}}$). This trend casts doubts on the use of HCN as an unbiased quantitative tracer of the dense molecular gas content in LIRGs and ULIRGs. A plausible scenario explaining the observed trend implies that X-rays coming from an embedded AGN may play a dominant role in the chemistry of molecular gas at $L_{\rm{IR}} \geq 10^{12}\,L_{\sun}$. We discuss the implications of this result for the understanding of LIRGs, ULIRGs and high redshift gas-rich galaxies.  
\end{abstract}



\section{Introduction}

Since their discovery by the IRAS survey, luminous infrared galaxies (LIRGs, $L_{\rm{IR}} > 10^{11}\,L_{\sun}$) and their extreme counterparts, ultraluminous infrared galaxies (ULIRGs, $L_{\rm{IR}} > 10^{12}\,L_{\sun}$), have been studied at virtually all wavelengths \citep{Sanders96}. The infrared (IR) emission of LIRGs and ULIRGs is known to stem from dust reprocessing of higher frequency radiation. However the nature of the power source of the huge IR emission in these galaxies
is still under debate \citep[][]{Sanders88,Veilleux95,Genzel98,Imanishi06,Risaliti06}. 

LIRGs and ULIRGs are known to possess large amounts of molecular gas as derived from CO(1--0) observations \citetext{\citealp{Sanders91,Solomon97}; Graci\'{a}-Carpio et al., in prep.}. The bulk of the molecular gas reservoir of ULIRGs lies in their central kpc \citep{Downes98}. \citet{Sanders88} first reported that the infrared--to--CO luminosity ratio in ULIRGs is anomalously high compared to that of normal galaxies and interpreted this result as evidence of the AGN power source scenario for ULIRGs. \defcitealias{Gao04a}{GS04ab}\citet[][hereafter GS04ab]{Gao04a, Gao04b} used HCN(1--0) observations to probe the dense molecular gas content of a sample of 65 nearby galaxies, including 25 LIRGs and 6 ULIRGs. Their results, showing a tight linear correlation between the IR and HCN luminosities over 3 orders of magnitude in $L_{\rm{IR}}$, were interpreted in terms of star formation as being the main power source in ULIRGs. However, doubts have been casted on the reliability of HCN as an unbiased tracer of dense molecular gas in LIRGs and ULIRGs. First, X-rays may significantly enhance HCN abundances in enshrouded AGNs \citep{Lepp96,Kohno01,Usero04}. Furthermore the excitation of HCN lines in LIRGs and ULIRGs might be affected by IR pumping through a 14\,\micron\ vibrational transition \citep{Aalto95}. Taken together, the possible caveats on the use of HCN observations call for the use of alternative tracers of dense gas in LIRGs and ULIRGs.

In this Letter, we present observations made with the IRAM \footnote{IRAM is supported by INSU/CNRS (France), MPG (Germany), and IGN (Spain).} 30m telescope in the 1--0 and 3--2 lines of HCO$^+$ of a sample of 16 galaxies including 10 LIRGs and 6 ULIRGs. The new HCO$^+$ spectra are complemented with CO(1--0) and HCN(1--0) observations either existing or acquired for this work. The HCO$^+$ and HCN J=1--0 lines have both similar critical densities and the bulk of their emission is expected to arise from dense molecular gas (n(H$_{2}$) $\geq 10^{4}$\,cm$^{-3}$). Preliminary results of this HCO$^+$ survey indicate that the HCN/HCO$^+$ luminosity ratio sharply increases with $L_{\rm{IR}}$ for LIRGs and ULIRGs.



\section{Observations}

The observations were carried out in November 2004 and August 2005 with the IRAM 30m telescope at Pico Veleta (Spain). 
We have obtained single-pointed spectra toward 16 objects with clear detections in all cases for the 1--0 line and with 10 detections for the 3--2 line of HCO$^{+}$. Our sample encompasses a $L_{\rm{IR}}$-range=10$^{11.3}$--10$^{12.5}\,L_{\sun}$. All targets are distant enough so as to recover the bulk of the emission from the molecular gas disks with the 30m telescope beam (FWHM $\sim$28\arcsec\,@\,89\,GHz and $\sim$9\arcsec\,@\,268\,GHz). HCN(1--0) and CO(1--0) single-dish data for the sample galaxies were taken from the literature \citetext{\citealp{Sanders91,Solomon92,Solomon97}; \citetalias{Gao04a}} or were observed by us with the 30m telescope (Table~\ref{table}). 
To compare the luminosity ratios derived for LIRGs and ULIRGs with those of {\it normal} galaxies ($L_{\rm{IR}} < 10^{11}\,L_{\sun}$), we have compiled a sample of 69 objects for which HCN, HCO$^{+}$ and CO data are available \citetext{\citealp{Nguyen92}; \citetalias{Gao04a}; this work}. 

\section{Caveats on the Use of HCN as Tracer of Dense Gas}

ULIRGs in \citetalias{Gao04a}'s sample have HCN--to--CO luminosity ratios up to $\sim$10 times higher than measured in normal galaxies. 
This result is corroborated by the new HCN observations of the sample of LIRGs and ULIRGs conducted for this work. As shown in Figure~\ref{dense-frac}\textit{a}, the HCN(1--0)/CO(1--0) luminosity ratio stays fairly constant around $\sim$0.03--0.04 for the normal galaxies 
of \citetalias{Gao04a}'s sample. Though with considerable scatter, the corresponding ratio for LIRGs and ULIRGs sharply increases, ranging from $\sim$0.05 to $\sim$0.30. However, if the HCO$^+$(1--0) line is adopted as an alternative tracer of the dense molecular gas, the situation changes significantly. As illustrated in Figure~\ref{dense-frac}\textit{b}, we find an average HCO$^{+}$(1--0)/CO(1--0) ratio close to $\sim$0.04 in normal galaxies using the sample of \citet{Nguyen92}. In the case of LIRGs and ULIRGs we do derive a comparatively higher HCO$^{+}$/CO ratio, suggestive of a correspondingly higher fraction of dense gas at higher $L_{\rm{IR}}$. However, the estimated HCO$^{+}$/CO ratio lies between $\sim$0.05 and $\sim$0.13, i.e., a factor of 2--3 less than the corresponding range derived for the HCN/CO ratio in LIRGs and ULIRGs. Further evidence that the dense molecular gas fraction of LIRGs and ULIRGs does not increase dramatically with  $L_{\rm{IR}}$ is provided by Figure~\ref{dense-frac}\textit{c} which shows that the derived HCO$^{+}$(3--2)/CO(1--0) ratio is roughly constant as a function of $L_{\rm{IR}}$ for $L_{\rm{IR}} > 10^{11}\,L_{\sun}$ with an average value of 0.026. We emphasize that the 3--2 line of HCO$^+$ is a secure tracer of dense (and warm) molecular gas, as the critical density for this transition is $n_{\rm{crit}} > 10^{6}$\,cm$^{-3}$.




Figure~\ref{trend}\textit{a} represents the HCN(1--0)/HCO$^{+}$(1--0) luminosity ratio as a function of $L_{\rm{IR}}$ for the LIRGs and ULIRGs in our sample. In spite of the small sample, we find an intriguing correlation between the HCN/HCO$^{+}$ luminosity ratio and $L_{\rm{IR}}$. The HCN/HCO$^{+}$ ratio is seen to increase by a factor of 5 from LIRGs to ULIRGs in our sample. We discuss below various scenarios that can explain why HCN seems to be over-luminous with respect to HCO$^+$ in LIRGs and ULIRGs:

{\it Infrared Pumping of the HCN(1--0) Line}---It has been argued that the excitation of HCN(1--0) in molecular clouds could be enhanced by infrared pumping via a 14\,\micron\ vibrational transition near strong mid-infrared sources \citep{Aalto95}. LIRGs and ULIRGs are likely to develop this type of environment for molecular gas. To evaluate infrared pumping we should ideally know the intensity of the mid-IR field felt by the molecules in each source. This would require a number of assumptions to be made on the geometry of the problem 
and on the degree of clumpiness of molecular gas. 
Following an empirical approach, we have compared the ratio of HCN and HCO$^{+}$ luminosities with an estimate of the excess in the 12\,\micron\ IRAS band: the ratio between the 12\,\micron\ and 100\,\micron\ IRAS density fluxes f$_{12}$/f$_{100}$ (see Figure~\ref{trend}\textit{b}; the 12\,\micron\ IRAS band covers the range of the HCN vibrational transition at 14\,\micron).
The observed lack of correlation between HCN(1--0)/HCO$^{+}$(1--0) and f$_{12}$/f$_{100}$ argues against infrared pumping of HCN(1--0) (see also discussion in \citetalias{Gao04a}). However, caution is required when interpreting the f$_{12}$/f$_{100}$ ratio as a measure of the MIR excess in the case of heavily obscured LIRGs and ULIRGs due to the high extinction in the MIR range \citep[e.g.,][]{Soifer02,Imanishi06}. The absence of correlation can not be taken as a firm evidence against IR pumping. On the other hand, the HCO$^{+}$ line could also be affected by IR pumping via a 12\,\micron\ vibrational transition. Even more, the conditions for infrared pumping for HCN(1--0) and HCO$^+$(1--0) are similar in terms of the required intensity of the IR field \citep{Carroll81}, a result that 
argues against this scenario. 

{\it Chemical Enhancement of HCN in Star Forming Regions}---There is observational evidence that the abundance of HCN can be enhanced in the molecular gas closely associated with high-mass star forming regions \citep[e.g.,][]{Blake87}. In contrast with quiescent molecular clouds, the measured HCN(1--0)/HCO$^{+}$(1--0) luminosity ratio is 
high ($\geq$1.5--2) in galactic GMCs \citep{Turner77}. More recently \citet{Pirogov99} published a survey of 34 bright FIR sources in the outer Galaxy identified as embedded high-mass star forming cores. Their results confirm that the HCN(1--0)/HCO$^{+}$(1--0) luminosity ratio is high in these regions: we have estimated that the average HCN(1--0)/HCO$^{+}$(1--0) luminosity ratio is $\simeq$2.3 for their cores sample. The densest (n(H$_2$) $\geq 10^{6}$\,cm$^{-3}$) and hottest ($T_{K} \geq 100$\,K) phase of the molecular star forming complexes (the `hot core' phase) can be a privileged niche for HCN chemistry, favoured by the evaporation of grain mantles. If the number of hot cores in LIRGs and ULIRGs exceeded that in our Galaxy by at least an order of magnitude, the observed HCN(1--0)/HCO$^{+}$(1--0) luminosity ratio could be thus accounted by a pure star formation scenario. In this case the reliability of HCN as a straightforward tracer of dense molecular gas should be equally put on hold.

{\it Chemical Enhancement of HCN in X-ray Dominated Regions (XDR) of AGNs}---Molecular gas can be exposed to a strong X-ray irradiation close to the central engines of active galaxies. In contrast to UV-photons, hard X-rays are efficient at penetrating huge gas column densities out to A$_{v} =$ 100--1000 \citep{Lepp96, Maloney96}. There is founded evidence that the circumnuclear disks (CND) of some nearby Seyferts can become giant X-ray dominated regions \citep{Kohno01, Usero04}. In the case of ULIRGs, it has been proposed that buried AGNs may create extended XDR rather than develop Narrow Line Regions (NLR) around their central engines, making the identification of AGNs in the optical difficult \citep{Imanishi06}. In particular, the influence of X-rays can enhance the abundance of HCN relative to other tracers of the dense molecular gas such as HCO$^+$ \citep{Lepp96}. 

\citet{Kohno01} have proposed the use of a diagnostic tool to distinguish between `pure' AGNs and `composite' starbursts+AGNs in nearby Seyferts, by exploring the range of variation of the HCN(1--0)/HCO$^{+}$(1--0) and HCN(1--0)/CO(1--0) ratios. We have applied such diagnostic tool to our sample. As shown in Figure~\ref{trend}\textit{c}, LIRGs and ULIRGs mostly populate the location of the `composite' AGN family in the diagram. However, compared to nearby Seyferts, we find more LIRGs/ULIRGs in the transition zone between `composite' and `pure' AGNs \citep[see Figure~3 of ][]{Kohno05}. This is an indication that XDR chemistry could be at work in enshrouded AGN in these intermediate objects.
According to Figure~\ref{trend}\textit{c}, Mrk~231 and Mrk~273 happen to lie in the `pure' AGN area. Of particular note, these two ULIRGs present firm evidence for hosting an embedded AGN in other wavelengths (see Table~\ref{table}).  Contrary to \citet{Kohno01}, who derived their ratios from interferometer intensity maps of the CND of nearby Seyferts, the ratios shown in Figure~\ref{trend}\textit{c} for LIRGs and ULIRGs represent total disk luminosity ratios. 
Therefore the case for a giant XDR in Mrk~231 and Mrk~273 is more compelling than in nearby Seyferts as their ratios are here weighted by the total content of their molecular disks, including the contribution from the starburst and the AGN.  



\section{Is HCO$^{+}$ a True Tracer of Dense Gas in ULIRGs?}

Most of the theoretically founded doubts have been thus far casted on the interpretation
of the HCN/CO ratio as to provide a quantitative estimate of the fraction of dense molecular gas in ULIRGs. However, the intriguing trend shown in Figure 2a could be alternatively interpreted as due to a possible deficiency of HCO$^+$ relative to HCN in ULIRGs. In particular, it is tempting to try to make a link between a hypothetical deficiency of HCO$^+$ and the well known deficiency of the [C\,II] 158\,\micron\ line relative to $L_{\rm{FIR}}$ in ULIRGs \citep{Luhman03}. Observational and theoretical evidence argue against this scenario, however. First, the bulk of the [C\,II] 158\,\micron\ and HCO$^+$(1--0) line emissions are not expected to arise, on average, from the same neutral phase of the ISM. Besides the fact that the critical densities of the two transitions markedly differ (n$_{crit}$\,HCO$^+$(1--0)\,$\gtrsim$\,10$\times$n$_{crit}$\,[C\,II]), PDR models predict that the bulk of the contribution to low-J HCO$^+$ lines should come from the inner layers of the PDR where the abundance of C$^+$ is diminished \citep{Sternberg95}. Moreover, significant non-PDR contributions to the HCO$^+$(1--0) line are very likely. In addition, most of the explanations brought forward thus far to solve the [C\,II] puzzle in ULIRGs invoke a quenching of the 158\,\micron\ line of C$^+$ rather than a lower ionization degree in the PDR phase anyway \citep{Luhman03}. In conclusion, there is no well founded evidence pointing to a physical link between a hypothetical deficiency of HCO$^+$ in ULIRGs and the well-established deficiency of [C\,II] 158\,\micron\ line in these galaxies. Other possible scenarios invoking a deficiency of HCO$^+$ should be explored in the future, however.

\section{Conclusions and Perspectives}

\begin{figure}
\centering
\epsscale{0.6}
\plotone{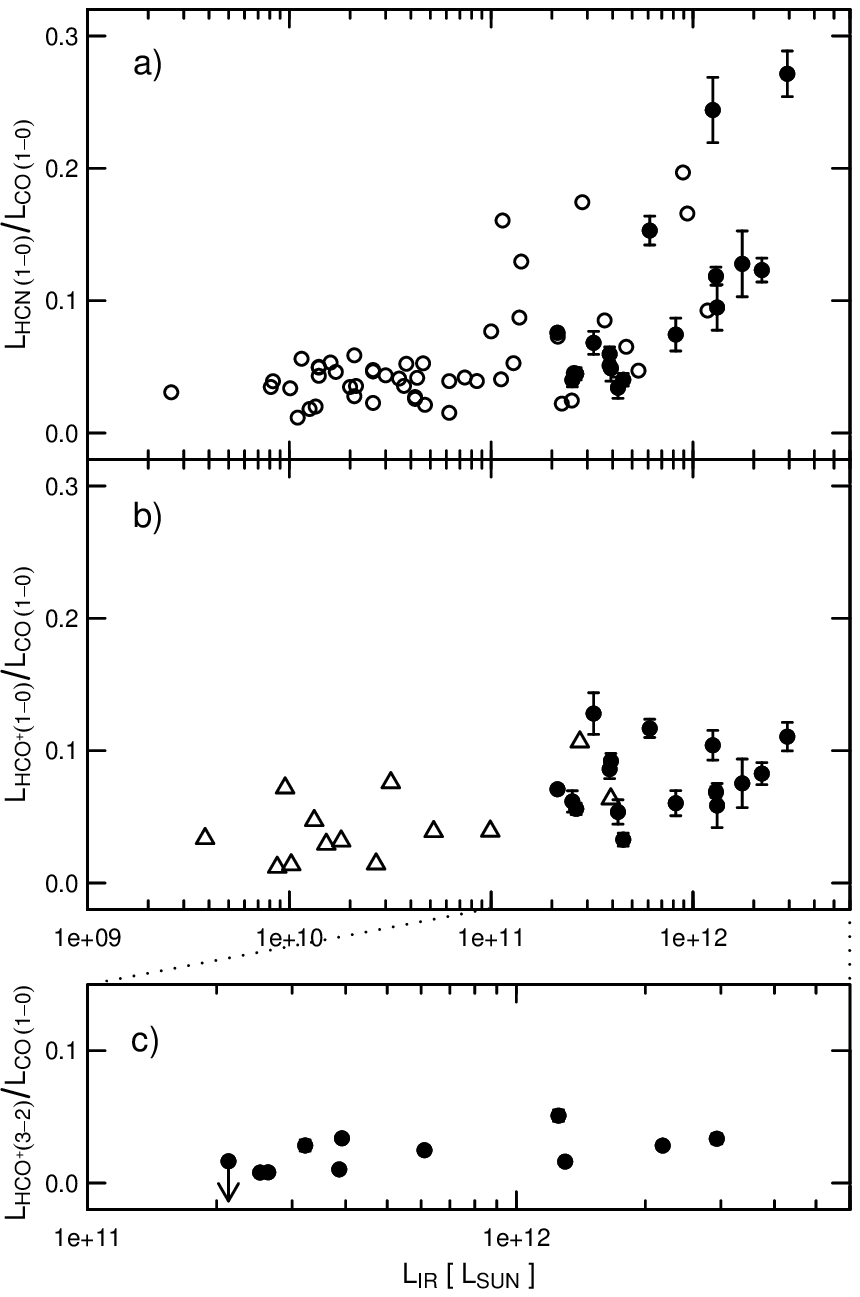}
\caption{Variation of the HCN(1--0)/CO(1--0) ({\bf a}) and HCO$^+$(1--0)/CO(1--0) ({\bf b}) luminosity ratios from normal galaxies ($L_{\rm{IR}} < 10^{11}\,L_{\sun}$) to LIRGs and ULIRGs ($L_{\rm{IR}} > 10^{11}\,L_{\sun}$). The open circles (triangles) correspond to galaxies from \citetalias{Gao04a}'s \citep{Nguyen92} sample. The HCO$^+$(3--2)/CO(1--0) ratio ({\bf c}) shows little variation within the explored range for LIRGs and ULIRGs. Filled circles correspond to our sample in all panels.}
\label{dense-frac}
\end{figure}

\begin{figure}
\centering
\epsscale{1}
\plotone{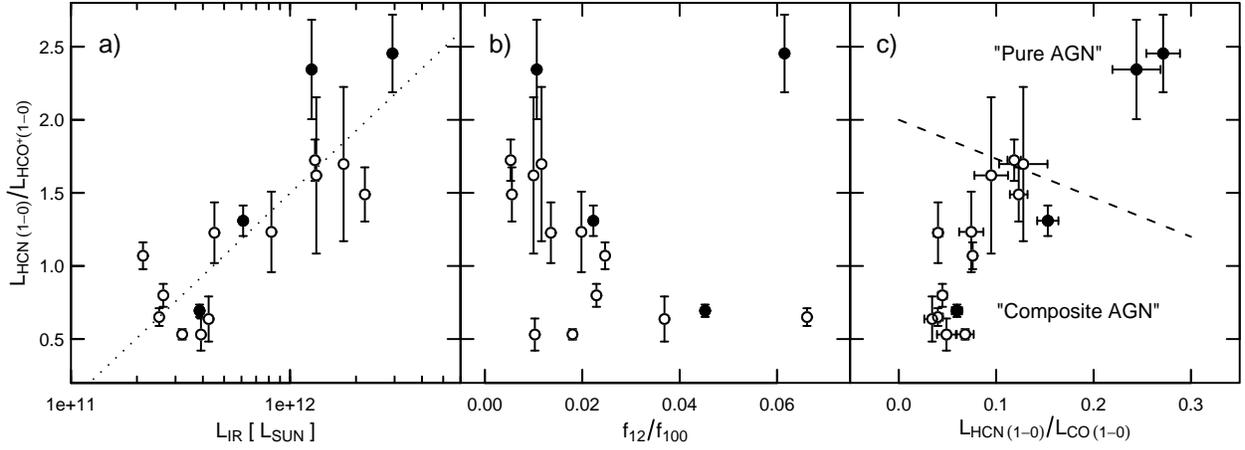}
\caption{{\bf a)} The correlation between the HCN(1--0)/HCO$^{+}$(1--0) luminosity ratio and $L_{\rm{IR}}$ is displayed for the sample of LIRGs and ULIRGs considered in this paper (the linear regression fit to the data gives a correlation coefficient $R = 0.84$). {\bf b)} The HCN(1--0)/HCO$^{+}$(1--0) ratio shows no correlation with the infrared colour f$_{12}$/f$_{100}$. {\bf c)} The location of LIRGs and ULIRGs of our sample in the 2D diagnostic diagram of \citet{Kohno01}. 
Filled symbols represent LIRGs and ULIRGs with secure identification of an embedded AGN.}
\label{trend}
\end{figure}

Preliminary results of the first HCO$^+$ survey conducted in LIRGs and ULIRGs provide evidence that HCN may not be a {\it true} tracer of dense molecular gas in these galaxies. Different mechanisms 
can make for HCN(1--0) being over-luminous with respect to HCO$^+$(1--0). The caveats on the interpretation of HCN observations highlight the need of surveys in other molecular species that together provide an unbiased estimate of the dense molecular gas fraction of LIRGs and ULIRGs. This question is central to disentangling the different power sources of the huge infrared luminosities of these galaxies. The variation of the HCO$^+$(1--0)/CO(1--0) ratio among the sample galaxies shows that the fraction of dense molecular gas of LIRGs and ULIRGs is, on average, a factor of $\sim$2 higher than that of normal galaxies (Figure~\ref{dense-frac}\textit{b}). This is considerably less than the corresponding number derived from HCN (i.e., a factor of $\sim$4--5). The increase in the dense gas fraction derived from HCO$^+$ would fall short of explaining the observed $L_{\rm{IR}}$ for ULIRGs in the purely star formation scenario of GS04ab. Instead, this result suggests that the contribution to $L_{\rm{IR}}$ from an embedded AGN source would amount to $\sim$50\,\% in the most extreme ULIRGs of our sample (Mrk~231 and Mrk~273). Most remarkably, the correlation found between the HCN(1--0)/HCO$^+$(1--0) ratio and  $L_{\rm{IR}}$ for LIRGs and ULIRGs runs in parallel with the long known general tendency of finding more AGN signatures in ULIRGs with increasing $L_{\rm{IR}}$ \citep{Veilleux95}. This seems to be confirmed by the derived location of Mrk~231 and Mrk~273 in the correlation plot of Figure~\ref{trend}\textit{a}, but also by the position of these galaxies in the diagnostic tool of Figure~\ref{trend}\textit{c}. More recently, HCN observations of an ultraluminous quasar at $z = 3.9$ \citep{Wagg05} have shown that the HCN/CO abundance ratio could be increased due to the presence of an AGN that dominates the bulk of the IR emission.

The extension of this work to galaxies with higher $L_{\rm{IR}}$ will help to shed light on the relative contribution of star formation and AGN to the huge infrared luminosity of high-$z$ objects. In particular, future results coming out from the extension of this survey to other molecular lines will allow us to better constrain the physical and chemical status of the dense molecular gas in ULIRGs (Graci\'{a}-Carpio et al., in prep.).


\acknowledgments This work has been partially supported by the Spanish MEC and Feder funds under grant ESP2003-04957 and by SEPCT/MEC under grants AYA2003-07584 \& AYA2002-01055.



\clearpage


\clearpage

\clearpage


\begin{deluxetable}{lccccccccccc}
\setlength{\tabcolsep}{0.02in}
\tabletypesize{\scriptsize}
\tablecaption{Sample Properties and Observational Results\label{table}}
\tablewidth{0pt}
\tablecolumns{12}
\tablehead{\colhead{} & \colhead{$D_{L}$} & \colhead{$L_{\rm{IR}}$} & \colhead{$L_{\rm{CO(1-0)}}$} & \colhead{$L_{\rm{HCN(1-0)}}$} & \colhead{$L_{\rm{HCO}^{+}\rm{(1-0)}}$} & \colhead{$L_{\rm{HCO}^{+}\rm{(3-2)}}$} & \multicolumn{4}{c}{Nuclear Classification}  & \colhead{} \\ 
\colhead{Source} & Mpc & $10^{11}\,L_{\sun}$ & $10^{8}\,L'$ & $10^{8}\,L'$ & $10^{8}\,L'$ & $10^{8}\,L'$ & L-band & MIR & Opt. & X-ray & \colhead{Ref.}}
\startdata
  \objectname[IRAS F12540+5708]{Mrk 231}            & 170 & 29.3 &  \bf 74 &  \bf 20 &  8.2 &      2.5 &     AGN &     AGN & Sy1 &    AGN  & 1     \\
  \objectname[IRAS F17207-0014]{IRAS\,17208\,--\,0014} & 173 & 21.9 &     131 &      16 &   11 &      3.7 &     SB  &     SB  & HII &    SB   & 1     \\
  \objectname[IRAS F12112+0305]{IRAS\,12112\,+\,0305}  & 297 & 17.6 &     101 &      13 &  7.6 &  \nodata &     SB  &     SB  & LIN &    SB   & 1     \\
  \objectname[IRAS F23365+3604]{IRAS\,23365\,+\,3604}  & 261 & 13.2 &      69 &     6.6 &  4.1 &  \nodata & \nodata &     SB  & LIN &    SB   & 2,3,4 \\
  \objectname[IRAS F15327+2340]{Arp 220}            &  73 & 13.0 &  \bf 70 & \bf 8.3 &  4.8 &      1.1 &     SB  &     SB  & LIN &    AGN? & 1     \\
  \objectname[IRAS F13428+5608]{Mrk 273}            & 152 & 12.5 &  \bf 55 &  \bf 13 &  5.7 &      2.8 &     AGN &     AGN & Sy2 &    AGN  & 1     \\
  \objectname[IRAS F05081+7936]{VII Zw 31}          & 220 &  8.2 & \bf 114 & \bf 8.4 &  6.9 &  \nodata & \nodata & \nodata & HII & \nodata & 5     \\
  \objectname[IRAS F16504+0228]{NGC 6240}           &  98 &  6.1 &  \bf 82 &  \bf 13 &  9.6 &      2.0 &     AGN &     SB  & LIN &    AGN  & 1     \\
  \objectname[IRAS 09126+4432]{Arp 55}              & 161 &  4.5 & \bf 127 &     5.1 &  4.2 &  \nodata & \nodata & \nodata & HII & \nodata & 2     \\
  \objectname[IRAS F01484+2220]{NGC 695}            & 131 &  4.2 &  \bf 48 &     1.7 &  2.6 &  \nodata & \nodata &     SB  & HII & \nodata & 2,6   \\
  \objectname[IRAS F13182+3424]{Arp 193}            &  94 &  3.9 &      38 &     1.8 &  3.5 &      1.3 & \nodata &     SB  & LIN & \nodata & 2,6   \\
  \objectname[IRAS F23007+0836]{NGC 7469}           &  66 &  3.9 &  \bf 34 &     2.0 &  2.9 &     0.35 & \nodata &     SB  & Sy1 &    AGN  & 2,7   \\
  \objectname[IC 694]{Arp 299\,A}                   &  42 &  3.2 &  \bf 13 &    0.89 &  1.7 &     0.37 & \nodata &     SB  & HII &    AGN? & 5,7,8 \\
  \objectname[NGC 3690]{Arp 299\,B+C}                 &  42 &  2.5 &  \bf 14 &    0.54 & 0.84 &     0.11 & \nodata &     SB  & HII &    AGN  & 5,7,8 \\
  \objectname[IRAS F23488+2018]{Mrk 331}            &  72 &  2.6 &  \bf 42 &     1.9 &  2.4 &     0.34 & \nodata &     SB  & HII & \nodata & 2,6   \\
  \objectname[IRAS F23488+1949]{NGC 7771}           &  57 &  2.1 &  \bf 43 &     3.2 &  3.0 & $<$ 0.70 & \nodata &     SB  & HII & \nodata & 2,6      
\enddata
\tablecomments{The luminosity distance, $D_{L}$, was calculated assuming $H_{0} = 75$ Km\,s$^{-1}$ Mpc$^{-1}$ and $q_{0} = 0.5$. The luminosities were computed according to \citet{Sanders96} and \citet{Solomon97} formulas. $L'$ = K km\,s$^{-1}$ pc$^{2}$, in $T_{\rm{mb}}$ scale. Single-dish data taken from the literature are marked in boldface. References for the nuclear classifications: (1) \citealt[][and references therein]{Risaliti06}; (2) \citealt{Veilleux95}; (3) \citealt{Rigopoulou99}; (4) \citealt{Teng05}; (5) NED; (6) \citealt{Lu03}; (7) \citealt{Genzel98}; (8) \citealt{Ballo04}.
}
\end{deluxetable}


\end{document}